\documentclass[letterpaper]{jpconf}
\usepackage{graphicx}
\begin{document}
\title{Reliability of P mode event classification using contemporaneous BiSON and GOLF observations}
\author{R. Simoniello$^{1}$, W.J. Chaplin$^{2}$,Y.P. Elsworth$^{2}$, R.A. Garc\'ia$^{3}$\\}
\address{$^{1}$Instituto de Astrof\'\i sica de Canarias (IAC), 38205, La Laguna, Tenerife, Spain\\
$^{2}$School of Physics and Astronomy, University of Birmingham, Edgbaston, Birmingham B15 2TT\\
$^{3}$Laboratoire AIM, CEA/DSM-CNRS - U. Paris Diderot - DAPNIA/SAp, 91191 Gif-sur-Yvette Cedex, France}
\ead{rosy@iac.es,wjc@bison.ph.bham.ac.uk,ype@bison.ph.bham.ac.uk,rfgarcia@cea.fr}
\begin{abstract}
We carried out a comparison of the signals seen in contemporaneous BiSON and GOLF data sets. Both instruments perform Doppler shift velocity measurements in integrated sunlight, although BiSON perform measurements from the two wings of potassium absorption line and GOLF from one wing of the NaD1 line. Discrepancies between the two datasets have been observed. We show,in fact, that the relative power depends on the wing in which GOLF data observes. During the blue wing period, the relative power is much higher than in BiSON datasets, while a good agreement has been observed during the red period.
\end{abstract}
\section{Introduction}
P-modes are standing acoustic waves in the solar interior. There are instances when it has been shown that solar flares are correlated with  modes of oscillation in the Sun just as the Earth is set ringing after a major earthquake. These events have been seen in high-degree modes (\cite{fog},\cite{kos}). We are interested in the possibility that the global modes are similarly stimulated and we seek to use the extensive BiSON dataset to search for these events. Large power events with strengths many times the mean level are seen and, moreover, the number is in excess of the predictions of stochastic excitation \cite{Cha1}. Our long-term aim is to categorize these rare events through their temporal features in both power and phase with the hope of  being able to distinguish natural from forced events. As part of this process of categorization of large excitations, we carried out a comparison of the signals seen in contemporaneous BiSON and GOLF data. The intention is to show that the power and the phase returned for the signal is in agreement for the two datasets. Given that, we can go on to use the very extensive BiSON data on its own. We show here that relative power in the two datasets depends on the wing in which the GOLF data are observed. We compare the observed power ratio to the predictions based on the heights in the atmosphere that the respective spectral lines are formed. We show also that the relative phases returned by the analysis are comparable.
\section{Data sets}
\subsection{Birmingham Solar Oscillation Network}
The Birmingham Solar Oscillation Network (BiSON) has been operating since 1976 and is one of the principal players in the field of seismology of global oscillation of the Sun. BiSON is a network of resonant scattering spectrometers (\cite{Cha}). It performs Doppler velocity measurements in integrated light from  the Fraunhofer absorption  line at 769nm. Only low degree modes with $\ell$=0,3 (azimuthal components $\ell$+m=even) are seen at high signal-to-noise. 
\subsection{Global Oscillation at Low Frequency}
The SOHO satellite is a project of international cooperation between ESA and NASA. It was launched on December  1995 (\cite{Dom},\cite{gab}). On board of this satellite are several instruments devoted to helioseismology. GOLF ($\bf G$lobal $\bf O$scillation at $\bf L$ow $\bf F$requency ) is a resonant scattering spectrometer that was designed to work at 4 points on the wings of the sodium absorption lines at 589.0 and 598.6nm. Due to the malfunctioning of the polarization system that switches between the two wings, GOLF has been operating in a single wing configuration. The nature of the signal has been demonstrated to be nearly pure velocity \cite{Pal}. We can distinguish three different observing periods:
a) from April 1996 to June 25 1998, GOLF operated  in the blue wings of the absorption lines;
b) after the SOHO vacation in September 1998 until to November 2002, it operated in the red wing configuration
c) from November 2002 to now it is in the blue wing configuration.
\subsection{Absorption line height formation and expected power in the modes}
The waves that are observed come primarily from the position in the atmosphere that corresponds roughly  to optical depth=2/3 for the wavelength of the absorption line used.  The difference in abundance of various elements in the Sun leads to their corresponding absorption lines being formed at different heights. As sodium is much more abundant in the Sun's atmosphere than potassium, the height formation of this line is much higher. 
It is shown that the height formation of BiSON data is relatively well localized at about 260 km above the base of the photosphere and does not vary much with the change in working point caused by the orbital velocity variation \cite{jim}. The GOLF working point sits higher up in the atmosphere. The median atmospheric height for the blue wing data is at 332 km, while for the red wing data it is at 480 km. However, there is a strong chromospheric component at 800 km for the red-wing data. This component is strongest when the line of sight velocity between the instrument and the Sun is at its maximum. 

The oscillations are confined within a cavity whose upper turning point lies below the height at which the absorption lines are formed. Hence, at the observation height the waves are evanescent. It can be shown (under some simplifying assumptions) that the energy density in the wave decreases exponentially the increasing height in the solar atmosphere with a scale height $H_{e}$. This last parameter is also a function of frequency ($\nu$) relative to the acoustic cut-off frequency ($\nu_{a}$)\cite{Chr}. The energy density is the product of the mass density and the square of the velocity amplitude (E=M$V^{2}$)and so we can predict how the observed velocity ($v$) will vary with height of formation of the observation line by comparison with a reference value $v_{o}$. The mass density is also decreasing exponentially with height in the atmosphere. The energy density scale height and the pressure scale height $H_{p}$ are connected by the following equation:  
\begin{equation}
H_{e}=\frac{H_{p}}{\left(1-\left(\frac{\nu}{\nu_{0}}\right)^{2}\right)^\frac{1}{2}}
\end{equation}
Under the assumption that there is no energy dissipation, we can say:
\begin{equation}
\left(\frac{v}{v_{o}}\right)^2=exp\left[h\left(\frac{1}{H_{p}}-{\frac{1}{H_e}}\right)\right]
\end{equation}
In the photosphere, the pressure scale height ($H_{p}$) is predicted by Vernazza to be approximately 120km \cite{Ver}. 
Hence the prediction is that signals formed higher in the atmosphere will be stronger than those formed lower down and the size of the effect will increase with increasing frequency when the observations are made close to the acoustic cut-off frequency. Between 3mHz and 4mHz the expected increase in the relative mode power is roughly 15$\%$. At low frequencies, the variation of the relative signal strengths with operating point in the atmosphere is rather weak.
\section{ Data Analysis}
\subsection{ Choice of data to compare} 
We compared the 5-minute oscillation spectrum obtained by analyzing BiSON and GOLF data  in the blue and red period separately. To do this, first we take  BiSON data set that matches GOLF data, taking therefore the same length of observations starting on the 11th of April 1996 and ending on the 31st December 2004. The BiSON data were moving meaned with a 25-sample filter. This attenuates signals below 1mHz. Then we extract from this time series one year of data starting on 11th of April 1996, where GOLF data were obtained only in the blue wing configuration. Then we took one year of data during the red wing configuration. In both cases we compared the power spectra of GOLF and BiSON.
\begin{figure}
\begin{center}
\includegraphics[scale=0.60]{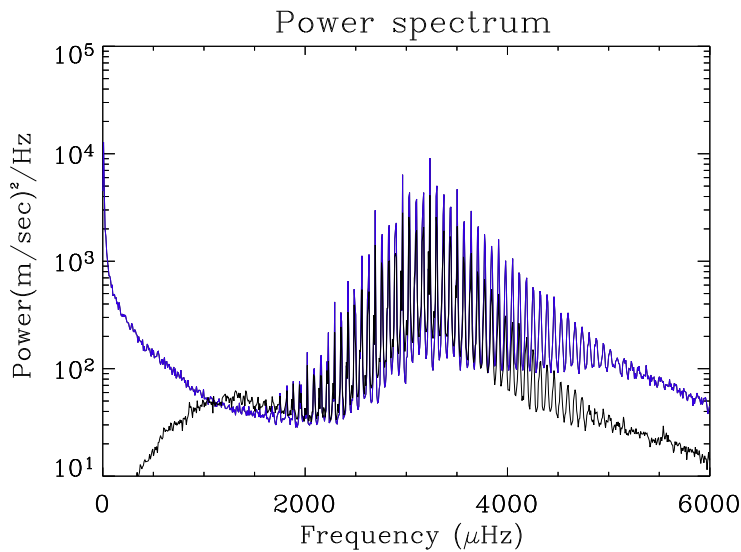}
\caption{Power spectrum comparison from GOLF blue wing configuration and BiSON dataset}
\label{fig:bluepowerspectra}
\end{center}
\end{figure}
\begin{figure}
\begin{center}
\includegraphics[scale=0.60]{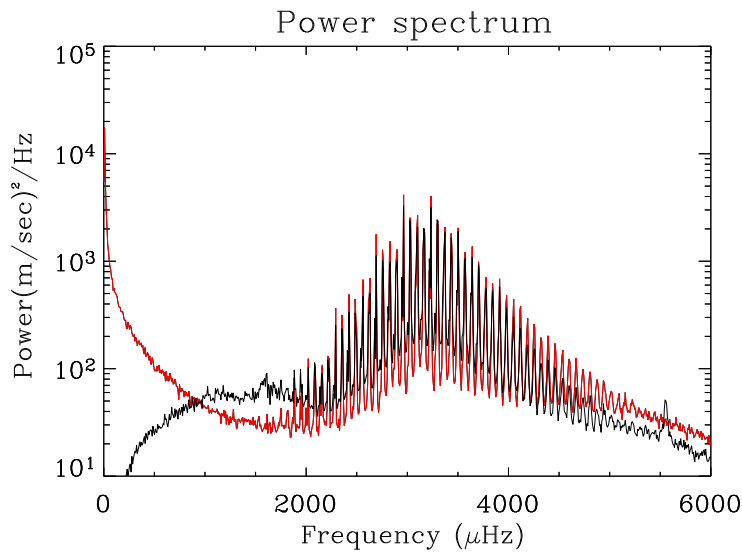}
\caption{Power spectrum comparison from GOLF red wing configuration and BiSON dataset}
\label{fig:redpowerspectra}
\end{center}
\end{figure}
Fig.~\ref{fig:bluepowerspectra} and Fig.~\ref{fig:redpowerspectra} show the comparison between BiSON and GOLF p-mode power spectra, where the color with which the GOLF data are shown, is indicative of the observing wing and the corresponding BiSON data are always shown in black. The expectation, based on the height of formation of the respective spectral lines, is that in the vicinity of 3mHz, the GOLF data will contain more power than the BiSON data and furthermore the red-wing data will be larger than the blue-wing data. As Fig.1-2 show, the situation is more complicated. 
First consider the blue wing data. In this case the line is formed at a height where the granulation noise in intensity is significant. The nature of the one-wing observation is that the instrument is more sensitive to this noise source than is the two-wing BiSON instrument. Hence, we believe that there is extra signal at low frequencies in the GOLF data by comparison with BiSON. We now turn to the red-wing data. Here the expectation (based on formation height) is that at 3 mHz power in the GOLF signal should be about twice the level of the BiSON signal. 
Fig.1-2 show that the observations do not entirely match the theoretical expectations. We believe that the disagreement is due to the known complications of scaling the GOLF data, and the extra granulation signal in the blue wing that will make it difficult to obtain an accurate comparison of contemporaneous data. So we chose to use  the red-wing data where there is expected to be much less granulation signal.
\subsection{ Power and phase temporal evolution features}
In order to study the temporal evolution of the modes we used sine-wave fitting on short datasets. The length of dataset chosen is half a day which corresponds to 1080 data points with 40s sample time. The resolution of such a spectrum is about 23 $\mu$Hz, which is enough to separate the odd and even $\ell$ modes but not enough to separate the individual components of the pair. For this reason we concentrate on $\ell$=1 modes for  which mode pair , the $\ell$=3 component, is very small. Under this assumption we can state that the power of the mode is located in the bin and therefore the height is essentially the power in the mode.
 There are several issues that we wish to address here. First we show how well the power in the GOLF and BiSON datasets track each other. 
\begin{figure}
\centering
\includegraphics[width=0.70\textwidth]{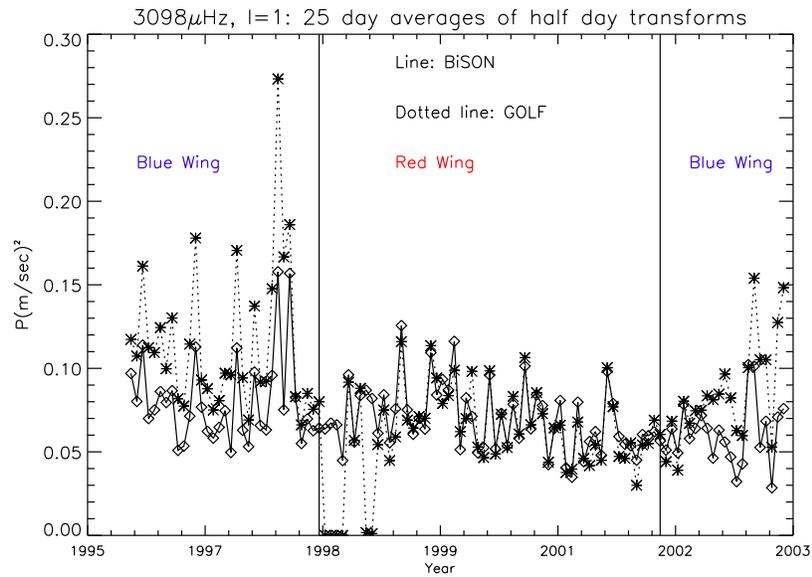}
\caption{GOLF-BiSON p-mode power temporal evolution} 
\label{fig:bisongolf}
\end{figure}
Fig.~\ref{fig:bisongolf} shows the temporal evolution of the mode at 3098 $\mu$Hz through the blue and the red period. It is clear that very similar signals are seen in the red period  from both sources and that the blue wing is more variable. We note in passing that the BiSON data do not always have 100$\%$ data fill in any one half-day period. This does not appear to invalidate the extraction of the power. We do choose to reject data with a fill less than 30$\%$ as the resolution is significantly degraded by the limited fill. Fig.~\ref{fig:corr} shows the correlation between red-wing period data for all $\ell$=1 modes from 2228 $\mu$Hz to 3355 $\mu$Hz. The correlation is good.
\begin{figure}
\begin{center}
\includegraphics[width=0.80\textwidth]{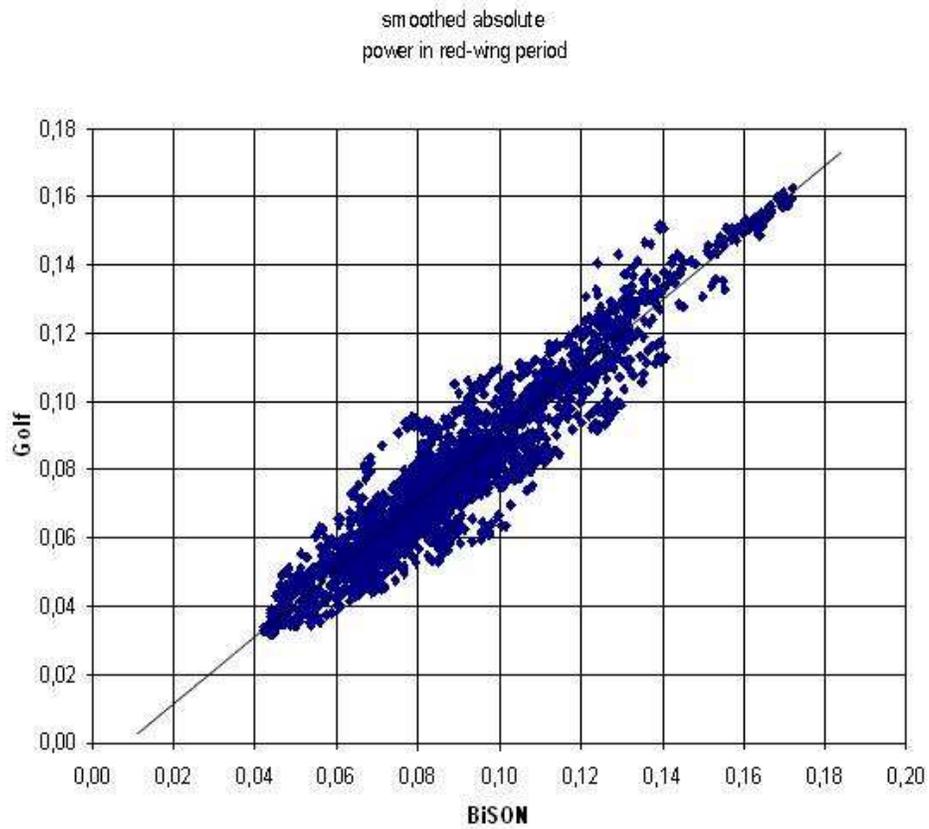}
\caption{Correlation between BiSON and GOLF blue wing configuration} 
\label{fig:corr}
\end{center}
\end{figure}
Having shown general agreement, we now look at the large excitations that will be the subject of further study. Fig.5 shows the temporal evolution of a large excitation in November 1998 and the agreement between GOLF and BiSON is good even in the case where the observed mode power is many times the average for the mode. This is strong evidence that the observed signal is solar in origin.Finally we consider the phase evolution of the modes. In order to characterize the excitation we plan to use both the power and the phase evolution. Fig.6 shows the phase for another of the large excitations in June 2002. The phase agreement is less good when the signal levels are very low and the background noise is more important. If we restrict ourselves to the situation where the mode power is greater than the average for the mode, the phase agreement is typically within about 5 degrees.
\begin{figure}
\begin{center}
\includegraphics[scale=0.50]{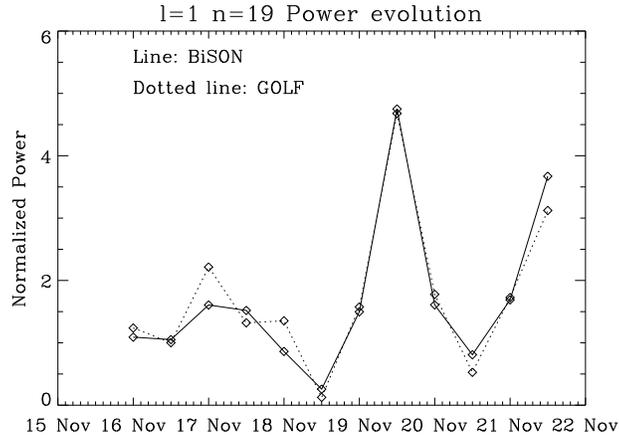}
\caption{P-mode power evolution over a week centered on 19th of November 1998}
\label{fig:powevolution}
\end{center}
\end{figure}
\begin{figure}
\begin{center}
\includegraphics[scale=0.50]{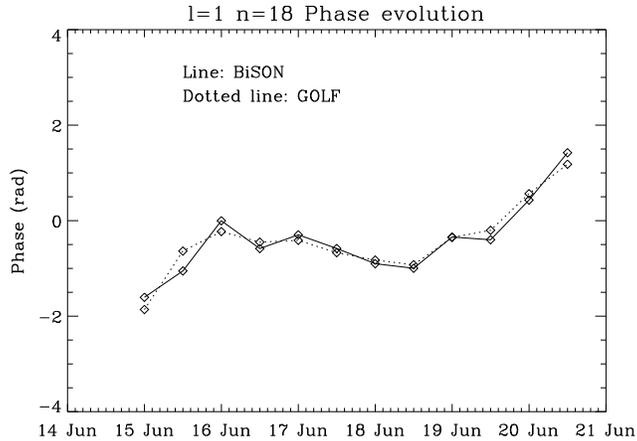}    
\caption{P-mode phase evolution over a week centered on 18th of June 2002}
\label{fig:phasevolution}
\end{center}
\end{figure}
\section{Conclusions}
We have used the comparison of contemporaneous GOLF and BiSON data to demonstrate that the power and phase evolution of the solar oscillations data are consistent between the two datasets and are therefore likely to be true characteristics of the oscillations of the Sun and not instrumental in origin. The red-wing GOLF data were found to be most useful for this.  The BiSON data were shown not to be compromised by their lack of perfect fill nor their ground-based observations. This opens up the possibility to use the full BiSON dataset spanning  three solar cycles in duration for further study of the large excitations and their links with solar activity. The next step in this project is to continue with the categorization of events with the eventual hope of being able to identify those that are stimulated by activity on the Sun.

\end{document}